\documentclass[aps,amsfonts,amsmath,nofootinbib,showpacs,byrevtex,preprintnumbers,showkeys,
prd,a4paper]{revtex4} 

\newcommand*{\be}{\begin{equation}}
\newcommand*{\ee}{\end{equation}}

\renewcommand*{\vec}[1]{\boldsymbol{\mathrm{#1}}}
\newcommand*{\uvec}[1]{\Hat{\boldsymbol{\mathrm{#1}}}}

\newcommand*{\vk}{\vec{k}}
\newcommand*{\uvk}{\uvec{k}}

\newcommand*{\bra}[1]{\langle #1 \rvert \mkern2mu}
\newcommand*{\ket}[1]{\mkern2mu \lvert #1 \rangle}

\DeclareMathOperator{\Det}{Det}
\DeclareMathOperator{\tr}{tr}
\DeclareMathOperator{\Tr}{Tr}

\newcommand*{\comm}[2]{ \bigl[ #1 , #2 \bigr]  }

\renewcommand*{\d}[1][]{\mathrm{d}^{#1}}

\newcommand{\e}{\ensuremath{\mathrm{e}}}

\newcommand*{\calO}{\ensuremath{\mathcal{O}}}

\newcommand{\rk}{\right)}
\newcommand{\lk}{\left(}

\begin{document}

\title{Heat-kernel expansion and counterterms of the Faddeev--Popov
determinant in Coulomb and Landau gauge}
\date{\today}

\author{Hugo Reinhardt}
\author{Davide R. Campagnari}

\affiliation{Institut f\"ur Theoretische Physik,
Universit\"at T\"ubingen,
Auf der Morgenstelle 14,
D-72076 T\"ubingen,
Germany}

\begin{abstract}
The Faddeev--Popov determinant of Landau gauge in $d$ dimensions and
Coulomb gauge in $d+1$ dimensions is calculated in the heat-kernel
expansion up to next-to-leading order. The UV-divergent parts in
$d=3,4$ are isolated and the counterterms required for a
non-perturbative treatment of the Faddeev--Popov determinant are
determined.
\end{abstract}

\maketitle


\section{Introduction}

In recent years there has been a strong interest in non-perturbative approaches
to continuum Yang--Mills theory and much research has been carried out in this
field. Most of these approaches rely on gauge fixing by the Faddeev--Popov
method. This concerns, in particular, the Dyson--Schwinger equation approach in
Landau gauge (for recent reviews see \cite{R1,R1a}) and in Coulomb gauge
\cite{R4a,R4}, and the Hamiltonian approach in Coulomb gauge \cite{R2,R3}.
Furthermore, in Coulomb and Landau gauge the Faddeev--Popov determinant is
assumed to dominate the infrared sector of the theory, see e.g.\ Ref.~\cite{R7}.

The standard way to renormalize the theory is to represent the
Faddeev--Popov determinant by ghost fields and use the BRST invariance of the
resulting local field theory. In some cases it is, however, advantageous not to
introduce ghost fields but to keep the Faddeev--Popov determinant explicit
and treat it non-perturbatively. This
is, in particular, the case in the Hamiltonian approach in Coulomb gauge \cite{R3}. The
renormalization of this theory requires then to identify the ultraviolet
singular pieces of the Faddeev--Popov determinant and to remove them by appropriate
counterterms. In Ref.\ \cite{R5} the UV counterterms of the Faddeev--Popov
determinant in $d = 3 + 1$ Coulomb gauge were identified using an approximate
expression for the Faddeev-Popov determinant \cite{R6}
\be
\label{F1-G1}
\ln\Det (- D \partial) = - \int \d[3]x \, \d[3]y \: A(x) \, \chi(x,y) \, A(y) \: ,
\ee
where $\chi$ is the so-called curvature, which represents the ghost loop
contribution to the gluon self-energy. The representation (\ref{F1-G1}) can be shown
to be correct to 2-loop order for the energy. Not surprisingly the required
counterterm to $\ln\Det(- D \partial)$ was found from this representation to be given by
a mass-like term
\be
\label{F1-G2}
C \int \d[3]x \: A^2 (x)
\ee
with the coefficient $C$ being linear in the UV-cutoff. In the
present paper we will calculate the counterterm to the Faddeev--Popov
determinant exactly without resorting to the representation (\ref{F1-G1}) and
also determine the precise numerical coefficient of the counterterm. For this
purpose we carry out a heat-kernel expansion of the Faddeev--Popov determinant in
$d = 3,4$ Landau gauge. The $d = 3$ Landau gauge corresponds to the $d = 3 + 1$
Coulomb gauge, the case of most interest to us.

The organization of the paper is as follows: In Sec.\ \ref{SEC2} we briefly
summarize the heat-kernel expansion of functional determinants.
In Sec.\ \ref{SEC3} we carry out the heat-kernel expansion for the
Fadeev--Popov operator in Landau gauge. The corresponding heat coefficients are
evaluated in Sec.\ \ref{SEC4} and the UV-divergent terms are isolated in Sec.\ \ref{SEC5}.
Finally in Sec.\ \ref{SUMMARY} we present our conclusions.


\section{\protect\label{SEC2}Heat kernel expansion of functional determinants}

We essentially follow Ref.~\cite{R9} here. The determinant of a positive
definite operator $M$ can be represented by the following proper-time integral
\be
\label{G1}
\Tr\ln M = - \int^\infty_{1/\Lambda^2} \d\tau \: \tau^{- 1} \: \Tr K (\tau) \, ,
\ee
where $\Lambda$ is an ultraviolet momentum cutoff and
\be
\label{G2}
K (\tau) = \e^{- M \tau}
\ee
is the heat-kernel. Furthermore, $\Tr$ includes both the trace over the
discrete indices $\tr$ and the integration over space time $\int \d[d]x$.
The heat-kernel (\ref{G2}) satisfies the heat equation
\be
\label{G3}
\partial_\tau K (\tau) + M K (\tau) = 0
\ee
and the boundary condition
\be
\label{G4}
K (\tau = 0) = 1 \, .
\ee
In the usual situation the operator under interest $M$ contains a ``free'' part
$M_0$, which can be treated exactly. In this case it is convenient to express
the full heat-kernel as
\be
\label{G5}
\bra{x} K (\tau) \ket{y} = \bra{x} K_0 (\tau) \ket{y} \: \bra{x} H (\tau) \ket{y} \, ,
\ee
where
\be
\label{G6}
K_0 (\tau) = \e^{- \tau M_0}
\ee
is the free heat-kernel satisfying
\be
\label{G7}
\lk \partial_\tau + M_0 \rk K_0 (\tau)  =  0 \, , \quad
K_0 (\tau = 0)  =  1 \,  ,
\ee
and $H (\tau)$ embodies all the ``interactions''. 
From the boundary conditions to the heat-kernels follows the boundary condition
for the interaction part
\be
\label{20}
H (\tau = 0) = 1 \, .
\ee
The interaction part $H (\tau)$ 
is then expanded in powers of the proper time $\tau$
\be
\label{G8}
\bra{x} H (\tau) \ket{y} = \sum^\infty_{n = 0} h_n (x, y) \: \tau^n \, ,
\ee
where the $h_n (x, y)$ are referred to as heat coefficients. Eq. (\ref{20}) 
implies
\be
\label{21}
h_0 (x, x) = 1 \, .
\ee
With (\ref{G5}) and (\ref{G8}) we find for the functional determinant
(\ref{G1})
\be
\label{G9}
\Tr \ln M = - \sum^\infty_{n = 0} \int^\infty_{1 / \Lambda^2} \d\tau \: \tau^{n - 1} \:
\Tr \lk K_0 (\tau) h_n \rk \, .
\ee
The advantage of the heat-kernel expansion (\ref{G9}) is that with increasing
order $n$ the terms become less and less ultraviolet singular and only
the first few terms are ultraviolet singular while, starting at a certain $n$
(which depends on the number of space-time dimensions), the terms are ultraviolet
finite. In the following we shall apply this heat-kernel expansion to the
Faddeev--Popov determinant in Landau gauge in $d = 3,4$ Euclidean dimensions.


\section{\protect\label{SEC3}Heat kernel evaluation of the Faddeev--Popov determinant in Landau
gauge}
 
The Faddev--Popov kernel in Landau gauge 
\be
\label{15}
\partial A = 0 
\ee
is given by
\be
\label{10}
M = - D \partial \, ,
\ee
where 
\be
\label{11}
D = \partial + A \, , \, A = A^a T^a \, , \, (T^a)^{b c} = f^{b a c}
\ee
denotes the covariant derivative in the adjoint representation of the gauge
field. Here and in the following all matrix valued quantities will be defined
in the adjoint representation of the gauge group. Furthermore, we
have absorbed the coupling constant in the gauge field. Note that the
Faddeev--Popov operator \eqref{10} is positive definite in the first Gribov
region, to which the gauge fields should be restricted, and hence its
determinant can be expressed by the proper-time integral \eqref{G1}. One
should mention, however, that the Faddeev--Popov operator has constant zero
modes corresponding to the global gauge invariance, which is not fixed by
the Landau gauge condition \eqref{15}. However, these zero modes occur also
for the ``unperturbed'' operator $M_0=-\partial^2$ and should hence cancel
in the ratio $\Det M / \Det M_0$, which we will consider here.

To illustrate the troubles of the Gribov region in the continuum theory,
consider constant gauge fields in $SU(2)$. For constant gauge fields the
eigenvalues of the Faddeev--Popov operator are given by (see Ref.~\cite{R10})
\be\label{11a}
\lambda_{\vk,\sigma} = \vk^2 - \sigma \, b \, |\vk| \, , \qquad \sigma = 0, \pm1
\ee
where
\be\label{11b}
b = \sqrt{(A_i^a \hat{k}_i)(A_j^a \hat{k}_j)} \, \qquad \uvk = \vk/|\vk| .
\ee
It is seen that even for very small $A_i^a=\mbox{const}$ there exists always
momenta $|\vk|<b$ so that the eigenvalue of Eq.~\eqref{11a} with $\sigma=1$
becomes negative and the constant field configuration is outside the first
Gribov region. This shows that in the continuum theory even arbitrary small
(e.g.\ constant) gauge fields are outside the first Gribov region. On the
lattice, constant gauge fields are inside the first Gribov region as long
as $b$ \eqref{11b} is smaller than the smallest non-zero momentum (the $\vk=0$
eigenvalue cancels against the $\vk=0$ eigenvalue of $M_0=-\partial^2$).
To guarantee that the sufficiently small gauge fields are inside the first
Gribov region we will introduce an IR-cutoff $\mu$ into the proper time integral
by inserting the factor $\exp(-\mu^2 \tau)$ into Eq.~\eqref{G9}.

Choosing the unperturbed part of the Faddeev--Popov operator as
\be
\label{12}
M_0 = - \partial^2  \, , 
\ee
the free heat-kernel (\ref{G6}) in $d$-Euclidean dimensions is given by
\be
\label{13}
\bra{x} K_0 (\tau) \ket{y} = (4 \pi \tau)^{- d/2} \:
\e^{- \frac{(x - y)^2}{4\tau}} \: . 
\ee
Once the unperturbed part $M_0$ is specified, a recursion relation for the
heat coefficients is derived from the heat equations \eqref{G3}, \eqref{G7}.
With the above choice of $M$ \eqref{10} and $M_0$ \eqref{12}, inserting the
ansatz (\ref{G5}) into the heat equation (\ref{G3}) and exploiting the
unperturbed heat equation (\ref{G7}) one obtains the following differential
equation for the interaction part
\be
\label{14}
\left[ \partial_\tau - \partial^2 + \frac{1}{\tau} (x - y) \partial +
\frac{1}{2 \tau} (x - y) A - A \partial \right] \bra{x} H (\tau) \ket{y}
= 0 \, .
\ee
Using the Landau gauge condition (\ref{15}) 
this equation can be expressed as
\be
\label{16}
\left[ \partial_\tau + \frac{1}{\tau} (x - y) D - \frac{1}{2 \tau} (x - y) A - D\partial \right]
\bra{x} H (\tau) \ket{y} = 0 \, .
\ee
Inserting here the expansion of the interaction kernel in terms of the proper
time (\ref{G8}) one derives the following recursion relation for the heat
coefficients $h_n (x, y)$
\be
\label{18}
\left[(k + 1) + (x - y) \lk D - \frac{1}{2} A \rk \right] h_{k + 1} (x, y) + \lk
- D \partial h_k (x, y) \rk = 0 \, 
\ee
with the initial condition
\be
\label{19}
(x - y) \lk D - \frac{1}{2} A \rk h_0 (x, y) = 0 \, .
\ee
Note that the equations (\ref{18}), (\ref{19}) are independent of the number of
dimensions so that the heat coefficients $h_k$ are the same in all dimensions.
The dependence on the number of dimensions is entirely contained in the free
heat-kernel (\ref{13}).

Due to the boundary condition (\ref{21}) the term with $n = 0$ yields the
determinant of the unperturbed kernel (\ref{12}). We therefore find with the
explicit expression for the unperturbed kernel (\ref{13}) from (\ref{G9})
\be
\label{22}
\ln \frac{\Det(- D \partial)}{\Det(- \partial^2)} =
- (4 \pi)^{-d/2} \sum^\infty_{n = 1} \: \int^\infty_{1/\Lambda^2}
\d\tau \: \e^{-\mu^2 \tau} \, \tau^{n - 1 - d/2} \: \Tr h_n \: ,
\ee
where
\be
\label{23}
\Tr h_n \equiv \int \d[d]x \: \tr h_n (x, x) \, .
\ee
From the representation (\ref{22}) it is seen that the terms with
\be
\label{24}
n - \frac{d}{2} \leq 0
\ee
are ultraviolet divergent, i.e.\ for $n < \frac{d}{2}$ we obtain power
divergences $\Lambda^{d - 2 n}$ while for $n = \frac{d}{2}$ we obtain a
logarithmic UV-divergence. Furthermore for $\mu=0$ the terms with $n - \frac{d}{2} \geq 0$
are infrared divergent, which is a manifestation of the presence of gauge
fields escaping the first Gribov region as discussed above. Keeping $\mu$
finite yields
\be
\label{4-27}
\ln \frac{\Det(- D \partial)}{\Det(- \partial^2)} = - \frac{1}{(4 \pi)^{d/2}}
\sum^\infty_{n = 1} \mu^{d - 2 n} \: \Gamma \lk n - \frac{d}{2} ,
\frac{\mu^2}{\Lambda^2} \rk \: \Tr h_n \, ,
\ee
where 
\be
\Gamma(a,z) = \int_z^\infty \d\tau \: \tau^{a - 1} \: \e^{- \tau}
\ee
is the incomplete Gamma function. In the UV-finite terms $n > \frac{d}{2}$ we
can let $\Lambda \to \infty$. Then the $\mu$-dependence of these terms is given
by
\be
\mu^{d - 2n}
\, ,
\ee
i.e.\ these terms are diverging for $\mu \to 0$. We are interested here only in
the UV-divergent part of the Faddeev--Popov determinant in $d = 3,4$, which can
be easily extracted. For this purpose it is sufficient to calculate the heat
coefficients up to including $h_2 (x, x)$, which we will do next.


\section{\protect\label{SEC4}Calculation of the heat coefficients}

The heat-kernels of generalized Laplacian operators were studied by a number
of authors, see e.g.\ Ref.~\cite{RX1}. Below we explicitly calculate the
leading and next-to-leading order heat coefficients of the Faddeev--Popov
operator \eqref{10}. To this purpose, it is useful to introduce
\be
\bar{D} = \partial + \frac12 \, A = D - \frac12 \, A \, ,
\ee
which is essentially a covariant derivative with a rescaled field. The 
Faddeev--Popov operator can then be re-expressed as
\be
-D \partial = -(\bar{D} + \tfrac12 A)(\bar{D} - \tfrac12 A) = - \bar{D}^2 + \tfrac14 A^2 ,
\ee
where we have the Landau gauge condition $\partial A = [\bar{D},A] = 0$,
and the recursion relation for the heat coefficients Eq.~\eqref{18} and
the initial condition Eq.~\eqref{19} become
\begin{subequations}
\begin{gather}
(x-y) \bar{D} h_0(x,y) = 0 \label{condition} , \\
\bigl[ (k+1) + (x-y) \bar{D} \bigr] h_{k+1}(x,y) =
\bigl[ \bar{D}^2 - \tfrac14 A^2 \bigr] h_k(x,y) \label{recursion} .
\end{gather}
\end{subequations}
Putting $k=0$ in the recursion relation \eqref{recursion} and taking the
coincidence limit $y \to x$ yields
\be\label{h1-1}
h_1(x,x) = \bigl[\bar{D}^2 h_0(x,y)\bigr]_{x=y} - \frac14 \, A^2 .
\ee
Acting with $\bar{D}_\mu \bar{D}_\nu$ on the initial condition
Eq.~\eqref{condition} yields
\begin{align}
0 &= ( \bar{D}_\mu \bar{D}_\nu + \bar{D}_\nu \bar{D}_\mu ) h_0(x,y) + \calO(x-y) \nonumber \\
&= \bigl\{ 2 \bar{D}_\mu \bar{D}_\nu + \comm{\bar{D}_\nu}{\bar{D}_\mu} \bigr\} h_0(x,y) + \calO(x-y)
\end{align}
In the coincidence limit this yields
\be\label{dd1}
\bigl[ \bar{D}_\mu \bar{D}_\nu h_0(x,y) \bigr]_{x=y} = \frac12 \, \bar{F}_{\mu\nu} \, ,
\ee
where $\bar{F}_{\mu\nu}$ is the field stregth tensor constructed with
the covariant derivative $\bar{D}$ (i.e.\ with the rescaled
gauge field $A/2$). This implies that the first term
in Eq.~\eqref{h1-1} vanishes, and the first heat coefficient reads
\be\label{h1}
h_1(x,x) = - \frac14 \: A^2 .
\ee
With the generators $T^a$ in the adjoint representation satisfying
$\tr T^a T^b = - N_C \, \delta^{ab}$ we eventually obtain
\be\label{h1res}
\Tr h_1 = \frac{N_C}{4} \int \d[d]x \, A_\mu^a(x) \, A_\mu^a(x) .
\ee

To find the second heat coefficient $h_2$ we put $k=1$ in the recusion
relation \eqref{recursion}, yielding in the coincidence limit
\be\label{h2-1}
2 \, h_2(x,x) = \left[ \bar{D}^2 - \frac14 A^2 \right] h_1(x,y) \bigr|_{x=y}
= \bigl( \bar{D}^2 h_1(x,y) \bigr)_{x=y} + \frac{1}{16} (A^2)^2 .
\ee
where we have used Eq.~\eqref{h1}. To calculate the right-hand side of Eq.~\eqref{h2-1}
we act with the operator $\bar{D}^2$ on the recursion relation \eqref{recursion}
with $k=0$
\be
\bar{D}^2 \bigl[ 1 + (x-y) \bar{D} \bigr] h_1(x,y) =
\bigl[ 3 \bar{D}^2 + \calO(x-y) \bigr] h_1(x,y)
\stackrel{\eqref{recursion}}{=} \bar{D}^2 \bigl[ \bar{D}^2 - \tfrac14 A^2 \bigr] h_0(x,y)
\ee
which yields from Eq.~\eqref{h2-1}
\be\label{h2-2}
h_2(x,x) = \frac16 \bigl( \bar{D}^2 \bar{D}^2 -
\frac14 \bar{D}^2 A^2 \bigr)h_0(x,y)\bigr|_{x=y} + \frac{1}{32} (A^2)^2 .
\ee
The second term in Eq.~\eqref{h2-2} can be rewritten as
\be
\bar{D}^2 A^2 = \comm{\bar{D}_\mu}{\comm{\bar{D}_\mu}{A^2}}
+ 2 \comm{\bar{D}_\mu}{A^2} \bar{D}_\mu + A^2 \bar{D}^2 .
\ee
The last two terms give a vanishing contribution when acting on $h_0$ in the
coincidence limit (see above), while the first one, being a total commutator,
gives no contribution when the trace in color space is taken. Thus the whole term
can be discarded. Applying the operator $\bar{D}^2 \bar{D}^2$
on the initial condition \eqref{condition} one gets
\be\label{h2-3}
0= \bar{D}^2 \bar{D}^2 \bigr[(x-y) \bar{D} h_0(x,y)\bigl] =
2 (\bar{D}^2 \bar{D}^2 + \bar{D}_\mu \bar{D}^2 \bar{D}_\mu) h_0(x,y) + \calO(x-y) .
\ee
Using the identity
\be
\bar{D}^2 \bar{D}^2 - \bar{D}_\mu \bar{D}^2 \bar{D}_\mu =
\bar{F}_{\mu\nu} \bar{F}_{\mu\nu} + \comm{\bar{D}_\mu}{\bar{F}_{\mu\nu}} \bar{D}_\nu
\ee
applied to $h_0$ and adding this to Eq.~\eqref{h2-3} we get finally
\be\label{h2-4b}
2 \bar{D}^2 \bar{D}^2 h_0(x,y) = \bigl( \bar{F}_{\mu\nu}^2 + \comm{\bar{D}_\mu}{\bar{F}_{\mu\nu}} \bar{D}_\nu \bigr) h_0(x,y).
\ee
The commutator term in Eq.~\eqref{h2-4b} gives no contribution in the coincidence
limit, since $\bar{D}_\mu h_0(x,y)=\calO(x-y)$ by the initial condition \eqref{condition}.
The second heat coefficient reads then
\be\label{h2-5}
\tr h_2(x,x) = \frac{1}{12} \tr \bar{F}_{\mu\nu} \bar{F}_{\mu\nu} + \frac{1}{32} \, \tr (A^2)^2 .
\ee
The first term in Eq.~\eqref{h2-5} can of course be expressed in terms of
the actual field strength tensor $F_{\mu\nu}$, which is related to $\bar{F}_{\mu\nu}$ by
\be
\bar{F}_{\mu\nu} = \frac12 \, F_{\mu\nu} - \frac14 \, \comm{A_\mu}{A_\nu} .
\ee
The heat coefficient \eqref{h2-5} then becomes
\be\label{h2}
\Tr h_2  =  \frac{1}{48} \int \d[d]x \: \tr \left[ \lk F_{\mu \nu} - \frac{1}{2} \comm{A_\mu}{A_\nu}
\rk^2 + \frac{3}{2} (A^2)^2 \right] \, .
\ee


\section{\protect\label{SEC5}The counterterms for $d = 3 + 1$
Coulomb gauge and $d = 4$ Landau gauge}

We are interested here in the divergent part of the Faddeev-Popov determinant in
$d = 3$ and $d = 4$  dimensions.  We start with the $d = 3$ dimensional case,
which corresponds to Coulomb gauge in $d = 3 + 1$ and is of most interest to us.
In this case the only UV-divergent term in
(\ref{22}) is $n = 1$. Since
\be
\Gamma \lk - \frac{1}{2}, x \rk = 2 \left[ \frac{e^{- x}}{\sqrt{x}} - \Gamma \lk
\frac{1}{2}, x \rk \right]
\ee
this term is IR-finite so that we can take the limit $\mu
= 0$ yielding
\be
\label{25}
\left.\ln\frac{\Det(- D \partial)}{\Det(- \partial^2 )}\right|_\mathit{div} = - \lk \frac{1}{4
\pi} \rk^{3/2} 2 \Lambda \: \Tr h_1 \, .
\ee
Inserting  the explicit expression for the heat-coefficient $h_1$ (\ref{h1res}),
we obtain for the divergent part of the Faddeev--Popov determinant
\be\label{fpcoulomb}
\left.\ln\frac{\Det(- D \partial)}{\Det(- \partial^2 )}\right|_\mathit{div} = - \frac{N_C}{2} \,
(4 \pi)^{-3/2} \Lambda \int \d[3]x \lk A^a_i (x) \rk^2 \, .
\ee
This result is consistent with Ref.\ \cite{R5}, where it was found that (to the order
considered) the renormalization of the Faddeev-Popov determinant requires a
counterterm of the type \eqref{F1-G2}. Here we have proven that, in general, in $d = 3$ there
is only one UV-divergent counterterm required and that the coefficient of this term
is linearly divergent.

In $d = 4$ the terms in (\ref{4-27}) with $n = 1, 2$ are UV-divergent yielding
\be\label{fplandau}
\left.\ln\frac{\Det(- D \partial)}{\Det(- \partial^2 )}\right|_\mathit{div} = - \frac{1}{(4 \pi)^2}
\left[ \Lambda^2 \, \Tr h_1 + \ln \frac{\Lambda^2}{\mu^2} \: \Tr h_2 \right] \, .
\ee
The heat coefficients $h_n$ are the same as in  $d = 3$ and are given by
Eqs.~(\ref{h1}) and (\ref{h2}).

A comment is here in order: It is well known that gauge invariance forbids
the occurrence of quadratically divergent mass terms. In fact, these terms
disappear due to a cancellation between contributions from ghost and gluon
loops. Since we are dealing here only with the Faddeev--Popov determinant,
i.e.\ with the ghost loop, such terms will occur.


\section{\protect\label{SUMMARY}Summary and conclusions}

We have performed a heat-kernel expansion of the Faddeev--Popov determinant
in Landau/Coulomb gauge. We have shown that in $d=3$ Landau ($d=3+1$ Coulomb)
gauge there is a single, unique UV-divergent counterterm required, which is
given by Eq.~\eqref{fpcoulomb}. It represents a mass term whose coefficient is
linearly UV-divergent. In $d=4$ Landau gauge there are two UV counterterms
required, which are respectively quadratically and logarithmically UV-divergent.
The latter term and all the UV-finite terms are IR-singular, reflecting the
absence of a mass scale in the original theory. However, as is well known the scale
anomaly gives rise to dimensional transmutation which manifests itself in
gluon condensation and the generation of a dynamical mass scale. To access the
UV-finite terms in the heat-kernel expansion, this dynamical mass scale has to
be included in the unperturbed heat-kernel. In principle, the heat-kernel
expansion is a type of gradient expansion, which should be applicable in
the infrared, provided the unperturbed kernel is adequately choosen. This
expansion is, however, bound to fail near the Gribov horizon, where the
Faddeev--Popov operator develops a zero eigenvalue. 


\begin{acknowledgments}
The authors would like to thank M.\ Quandt for a critical reading
of the manuscript. This work was supported by the Deutsche
Forschungsgemeinschaft (DFG) under contract No.~Re856/6-3 and
the Cusanuswerk--Bisch\"ofliche Studienf\"orderung.
\end{acknowledgments}



\begin{thebibliography}{99}
\bibitem{R1}
C.~S.~Fischer, J.\ Phys.\ \textbf{G32}, R253 (2006); [arXiv:hep-ph/0605173].
\bibitem{R1a}
D.~Binosi and J.~Papavassiliou, Phys.\ Rept.\  \textbf{479}, 1 (2009); [arXiv:0909.2536 [hep-ph]].
\bibitem{R4a}
D.~Zwanziger, Nucl.\ Phys.\ \textbf{B518}, 237 (1998).
\bibitem{R4}
P.~Watson and H.~Reinhardt, Phys.\ Rev.\ \textbf{D75}, 045021 (2007); [arXiv:hep-th/0612114].
\bibitem{R2}
A.~P.~Szczepaniak and E.~S.~Swanson, Phys.\ Rev.\ \textbf{D65}, 025012 (2002); [arXiv:hep-ph/0107078].
\bibitem{R3}
C.~Feuchter and H.~Reinhardt, Phys.\ Rev.\ \textbf{D70}, 105021 (2004); [arXiv:hep-th/0408236].
Phys.\ Rev.\ {\bf D76}, 125016 (2007); [arXiv:0709.0140 [hep-th]].
\bibitem{R7}
D.~Zwanziger, Phys.\ Rev.\ \textbf{D70}, 094034 (2004); [arXiv:hep-ph/0312254].
\bibitem{R5}
D.~Epple, H.~Reinhardt, W.~Schleifenbaum and A.~P.~Szczepaniak,
Phys.\ Rev.\ \textbf{D77}, 085007 (2008); [arXiv:0712.3694 [hep-th]].
\bibitem{R6}
H.~Reinhardt and C.~Feuchter, Phys.\ Rev.\ \textbf{D71}, 105002 (2005); [arXiv:hep-th/0408237].
\bibitem{R9}
D.~Ebert and H.~Reinhardt, Nucl.\ Phys.\  \textbf{B271}, 188 (1986).
\bibitem{R10}
H.~Reinhardt and W.~Schleifenbaum, Annals Phys.\  \textbf{324}, 735 (2009); [arXiv:0809.1764 [hep-th]].
\bibitem{RX1}
P.~B.~Gilkey, J.\ Diff.\ Geom.\ \textbf{10}, 601 (1975).
\end{thebibliography}
\end{document}